\documentclass{PoS}
\usepackage{amsmath}
\usepackage{subfigure}
\title{$\Delta I = 3/2$, $K\rightarrow \pi\pi$ Decays with a Nearly Physical Pion Mass}

\ShortTitle{$\Delta I = 3/2$, $K\rightarrow \pi\pi$ Decays with a Nearly Physical Pion Mass}

\author{{Elaine J Goode}\\%
	University of Southampton, School of Physics and Astronomy, Highfield, Southampton, SO17 1BJ, United Kingdom\\
        E-mail: \email{ejg4g08@soton.ac.uk}
}
\author{\speaker{Matthew Lightman}\\
        Department of Physics, Columbia University, New York, NY 10027 USA\\
	Department of Physics, Washington University, St. Louis, MO 63130 USA\\
        E-mail: \email{mlightman@physics.wustl.edu}}

\abstract{The $\Delta I = 3/2$ $K\rightarrow \pi\pi$ decay amplitude is calculated on RBC/UKQCD $32^3\times 64 $, $L_s=32$ dynamical lattices
with $2+1$ flavors of domain wall fermions using the DSDR and Iwasaki gauge action. The calculation is performed with a single pion mass ($m_{\pi}=141.9(2.3)$ MeV,
 partially quenched) and kaon mass ($m_K=507.4(8.5)$ MeV) which are nearly physical, and with nearly energy conserving kinematics. Antiperiodic boundary conditions in 
two spatial directions are used to give the two pions non-zero ground state momentum. 
Results for time separations of 20, 24, 28 and 32 between the kaon and two-pion sources are computed
and an error weighted average is performed to reduce the error. We find prelimenary results for Re$(A_2)=1.396(081)_{\text{stat}}(160)_{\text{sys}}\times10^{-8}$ GeV 
and Im$(A_2) = -8.46(45)_{\text{stat}}(1.95)_{\text{sys}} \times 10^{-13}$ GeV. } 

\FullConference{The XXVIII International Symposium on Lattice Field Theory, Lattice2010\\
		June 14-19, 2010\\
		Villasimius, Italy}

\begin{document}

\vspace{-0.05 in}
\section{Introduction}
\vspace{-0.05 in}

Precise lattice calculations of $K \rightarrow \pi\pi$ decays will provide quantitative insight into 
the origin of the $\Delta I = 1/2$ rule and direct CP violation in kaon decays. Previous calculations have relied on the quenched
approximation or have attempted to use chiral perturbation theory to extrapolate from heavy quark masses down to physical masses, or both 
\cite{kim,kim_christ,blum,CPPACS,chiral_extrap,Boucaud,Lightman_Goode}. 
The calculation presented here avoids both these sources of error by computing the 
$K\rightarrow \pi\pi$ amplitude directly using dynamical lattices with 2+1 flavours of domain wall fermions (DWF) at near physical
pion mass. We use RBC/UKQCD $32^3 \times 64$, $L_s=32$ lattices which use the Dislocation Suppressing Determinant Ratio (DSDR) plus Iwasaki gauge action with inverse lattice
spacing $a^{-1}\approx1.4$ GeV ($\beta=1.75$) and domain wall height $M_5=1.8$. We use ensembles generated with $am_l^{\mathrm{sea}} = 0.001$, 
$am_s^{\mathrm{sea}} = 0.045$, corresponding to a unitary pion mass of $m_{\pi} \approx 180$~MeV.

\vspace{-0.05 in}
\section{Four-Quark Operators and The Effective Hamiltonian}
\vspace{-0.05 in}

The weak interactions and the effects of heavier quarks can be included in the lattice QCD simulation by evaluating matrix elements of
an effective Hamiltonian \cite{ciuchini, buchala}. In particular the conventions of \cite{blum} are used. We calculate matrix elements of
four-quark operators between $K$ and $\pi\pi$ states. In this paper the amplitude $A_2$ is calculated, which requires the evaluation of 
matrix elements of three operators. These operators are classified by how they transform under $SU(3)_L \times SU(3)_R$: 
$Q_{(27,1)}$,  $Q_{(8,8)}$ and $Q_{(8,8)\text{mx}}$. Progress in calculating $A_0$ is described in \cite{qi}.

\vspace{-0.05 in}
\section{Boundary Conditions}
\vspace{-0.05 in}

We wish to simulate the $K\rightarrow \pi\pi$ decay at physical kinematics, which requires the final state pions to have a non-zero momentum.
This is achieved by imposing antiperiodic boundary conditions on the quark fields in one or more spatial directions. The allowed momenta of the quark are then
given by $  p_n = (\pi + 2\pi n)/L$, where $L$ is the spatial extent of the lattice. 

We relate the physical $\left\langle \pi^+ \pi^0 \right \vert Q^{\Delta I = 3/2}_{\Delta I_z=1/2} \left \vert K^+ \right \rangle$ matrix
element to the unphysical matrix element\\
$\left\langle \pi^+ \pi^+ \right \vert Q^{\Delta I = 3/2}_{\Delta I_z = 3/2} \left \vert K^+ \right \rangle$ using
the Wigner-Eckart theorem. 
This simplifies the operators and allows us to use periodic boundary conditions on the up- and strange-quarks while using
antiperiodic boundary conditions only on the down-quark, thus giving the two pions momentum
while the kaon remains at rest.

If antiperiodic boundary conditions are imposed on the d-quark in only the $x$ direction 
with periodic boundary conditions in the $y$ and $z$ directions then we can have a two-pion 
ground state in which one pion has momentum $p_x = \frac{\pi}{L}$ and the other pion has momentum $p_x = -\frac{\pi}{L}$. The antiperiodic boundary
conditions allow us to extract non-zero momentum pions without the need to fit to an excited state, which would have been necessary had we imposed 
periodic boundary conditions on all the quark fields. In principle we can impose antiperiodic boundary conditions on the d-quark in one, two, or all 
three spatial directions corresponding to individual ground state pion momenta of
$p=\pi/L,\, \sqrt{2}\pi/L$ and $\sqrt{3}\pi/L$ respectively.

\vspace{-0.05 in}
\section{Details of the Calculation}
\vspace{-0.05 in}

The calculation was carried out on 62 configurations of dynamical $32^3 \times 64$ lattices using DSDR+Iwasaki gauge action and domain wall 
fermions with $L_s=32$, generated on BG/P machines at Argonne National Laboratory. Further details of the 
ensemble generation are given in~ \cite{bob}. The inverse lattice spacing is $a^{-1}=1.365(22)\text{ GeV}$,
the physical volume is $(\text{4.62 fm})^3$ and we set the light and strange valence quark masses to $am_l = 0.0001$ and $am_s=0.049$ respectively.
This corresponds to a pion mass of $m_{\pi} = 141.9(2.3) \text{ MeV}$ and a kaon mass of $m_K = 507.4(8.5) \text{ MeV}$. 

We combined propagators with periodic and antiperodic boundary conditions in the time direction in order to double the effective time extent
of the lattice. The meson correlation functions contained propagators which were computed with a source at $t=0$ (corresponding to (P+A) bc) and $t=64$ 
(corresponding to (P-A) bc).
We also generated strange-quark propagators with sources at $t_K = $ 20, 24, 28, 32, 36, 40 and 44 in order to calculate 
$K\rightarrow \pi\pi$ correlators with kaon sources at these times, while the two-pion sources remained at either $t=0$ or $t=64$. 
Thus we could achieve time separations between the kaon and two pions of 20, 24, 28 and 32 in two different ways which doubled the statistics.
These separations were chosen so that the signals from the kaon and two pions did not decay to noise before reaching the four-quark operator $Q$.

For the kaon and pions with zero momentum we use propagators with Coulomb gauge-fixed wall sources. For the two pions with non-zero momentum
we use the same type of propagators for the u quark but used propagators with antiperodic spatial boundary conditions for the d-quark with
Coulomb gauge-fixed momentum wall sources of the ``cosine'' type
\begin{equation}
 s_{\mathbf{p},\text{ cos}}(\mathbf{x}) = \cos\left(p_x x\right) \cos\left(p_y y\right) \cos\left(p_z z\right).
\end{equation}

 We use the same cosine source for each d-quark, which introduces a cross term that couples to two-pion states with non-zero total 
momentum. For example, if we consider giving momentum in only the $x$ direction the product of the sources of the two d-quarks is
\begin{equation}
\begin{split}
 s_{\mathbf{p},\text{ cos}}(\mathbf{x}_1)s_{\mathbf{p},\text{ cos}}(\mathbf{x}_2) & = \cos\left(\frac{\pi}{L} x_1\right) \cos\left(\frac{\pi}{L}x_2\right) \\
& = \frac{1}{4}\left( e^{i\frac{\pi}{L}x_1}e^{i\frac{\pi}{L}x_2} + e^{i\frac{\pi}{L}x_1}e^{-i\frac{\pi}{L}x_2} 
+ e^{-i\frac{\pi}{L}x_1}e^{i\frac{\pi}{L}x_2} +e^{-i\frac{\pi}{L}x_1}e^{-i\frac{\pi}{L}x_2} \right).
\end{split}
\label{eq:multcos}
\end{equation}
We require the two pions to have individual momentum $\mathbf{p}_1 = \frac{\pi}{L}\mathbf{\hat x}$ and
  $\mathbf{p}_2 = -\frac{\pi}{L}\mathbf{\hat x}$, but the first and last terms of equation \eqref{eq:multcos} couple to two-pion states with 
total momentum $2\frac{\pi}{L}$ and $-2\frac{\pi}{L}$ respectively. We eliminate the unwanted terms in the two-pion correlator by using pure exponential
momentum sinks which constrain the final state to have zero total momentum. In the $K\rightarrow \pi\pi$ correlator, the zero
momentum kaon has a similar effect on the cosine sources of the two-pions. Had we used the more conventional momentum source 
\begin{equation}
 s_{\mathbf{p}}(\mathbf{x}) = e^{i\mathbf{p}\cdot\mathbf{x}}
\end{equation}
we would have needed to perform two separate d-quark inversions with momentum $+\mathbf{p}$ for one and $-\mathbf{p}$ for the other. The cosine 
source eliminates one of these inversions. In practice we only compute d-quark propagators with antiperiodic boundary conditions in $0$ or $2$ spatial directions,
corresponding to pions with ground state momenta $p=0$ and $p = \sqrt{2}\pi/L$. This choice is motivated by the expectation that for our choice of quark masses, $p=\sqrt{2}\pi/L$
will correspond to on-shell kinematics.

\vspace{-0.05 in}
\section{Analysis and Results}
\vspace{-0.05 in}

\begin{figure}[h]
 \centering
\subfigure[$(27,1)$ operator]{\includegraphics*[width=0.31\textwidth]{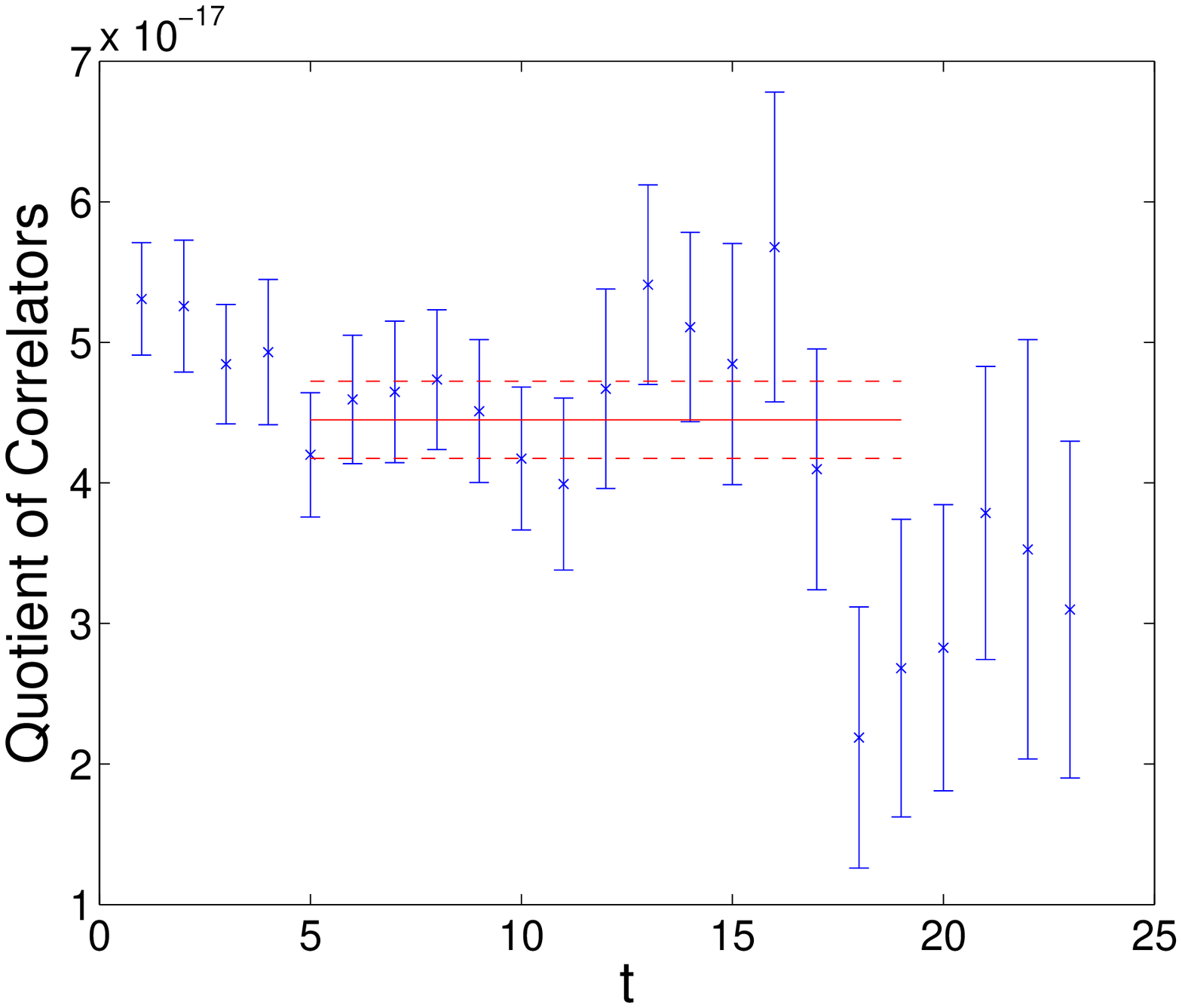}}
\subfigure[$(8,8)$ operator]{\includegraphics*[width=0.31\textwidth]{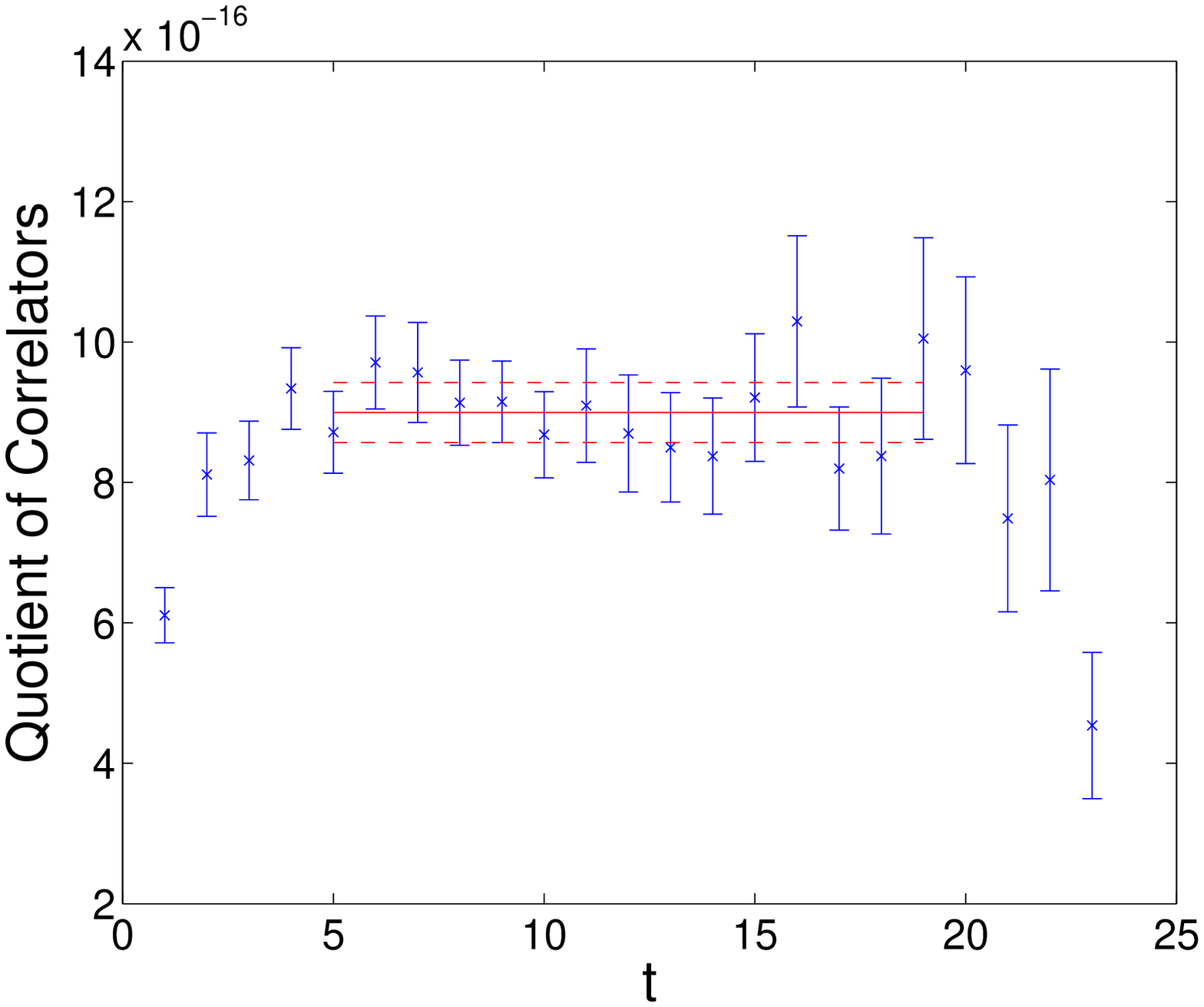}} 
\subfigure[$(8,8)_{\text{mx}}$ operator]{\includegraphics*[width=0.31\textwidth]{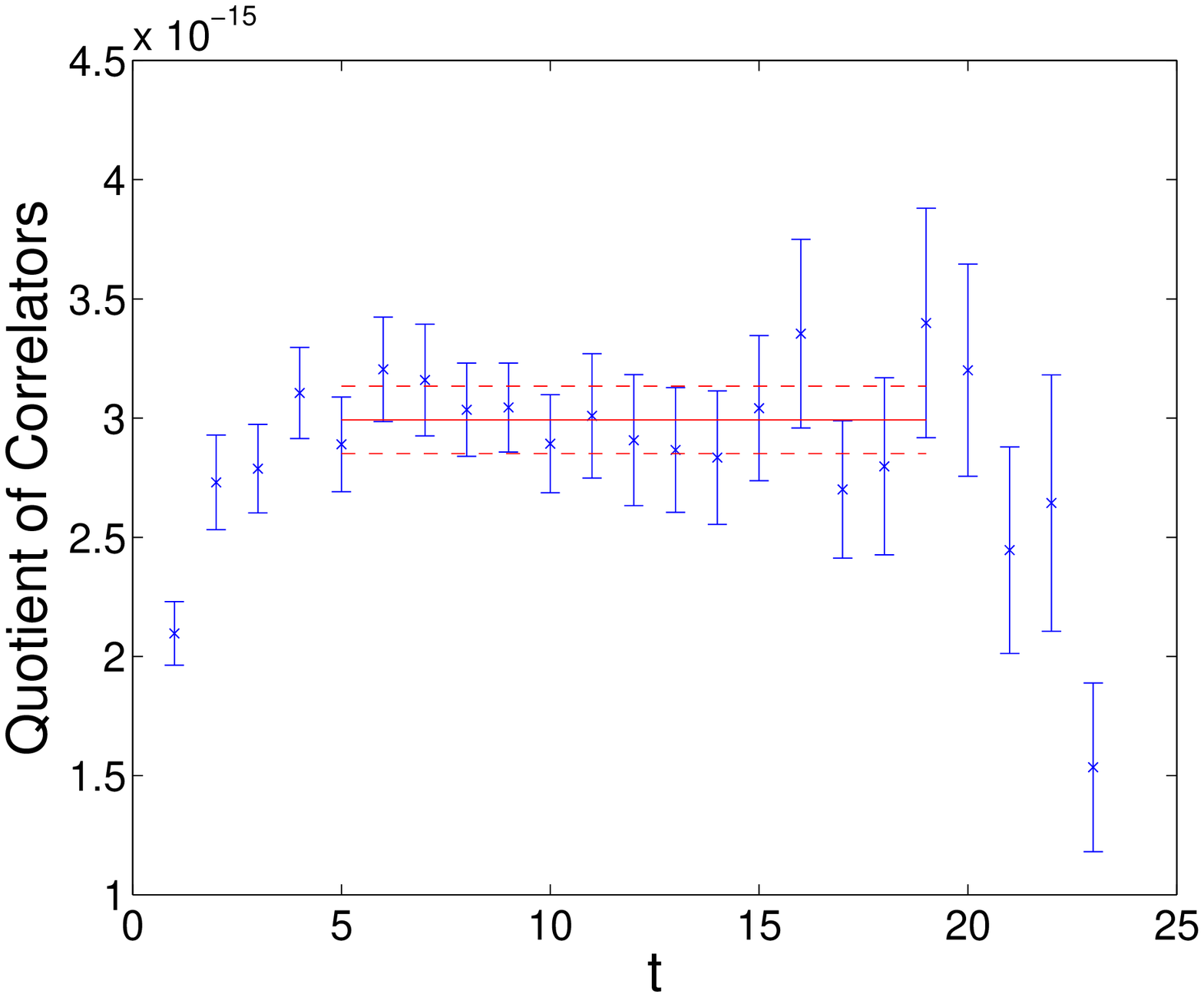}}
\caption{\label{fig:q_plot} $K\rightarrow \pi\pi$ quotient plots for $p= \sqrt{2}\pi/L$.
The two pion source is at $t=0$ while the kaon source is at $t=24$. The dashed line shows the error on the fit}
\end{figure}

We extract the $K \rightarrow \pi\pi$ matrix element $\mathcal{M}$ by fitting a constant to the left hand side of \eqref{eq:quot1}

\begin{equation}
 \frac{C^i_{K\pi\pi}(t)}{C_K(t_K-t)C_{\pi\pi}(t)} = \frac{\mathcal{M}_i}{Z_K Z_{\pi\pi}}.
\label{eq:quot1}
\end{equation}
$C^i_{K\pi\pi}$ is the $K \rightarrow \pi\pi$ correlator with a
kaon source at $t_K$, $i$ labels the four-quark operator $Q_i$ which is inserted at time $t$, and $Z_K$ and $Z_{\pi\pi}$ are calculated from the kaon and 
two-pion correlators respectively, whose sources are at $t=0$.
The left hand side of equation \eqref{eq:quot1} is plotted in figure \ref{fig:q_plot} for each of the three operators. The figure demonstrates that sufficiently far from 
the kaon and two-pion sources we are justified in fitting to a constant. The fit results for $\mathcal{M}_i/(Z_K Z_{\pi\pi})$ are indicated on the plot.

The finite volume matrix elements are related to the infinite volume amplitudes $A_i$ using the Lellouch-L\"uscher factor 
\cite{LLfactor, LLfactor_sachrajda}. In particular we have
\begin{equation}
 A_i = \left[ \frac{\sqrt{\nu/4}}{\pi q_{\pi}} \sqrt{\frac{\partial \phi}{\partial q_{\pi}} + \frac{\partial \delta}{\partial q_{\pi}} } \right]
 \frac{1}{\sqrt{\nu}} \sqrt{m_K} E_{\pi\pi} \mathcal{M}_i
\label{eq:amp}
\end{equation}
where the quantity in square brackets (denoted by LL in table \ref{tab:derivs}) 
contains the effects of the Lellouch-L\"uscher factor beyond the free field normalization. $E_{\pi\pi}$ is the energy of the 
two-pion state, $\delta$ is the s-wave phase shift, $\nu$ is a
 factor counting the free-field degenerate states, 
$q_{\pi}$ is a dimensionless quantity related to the individual pion momentum $k_{\pi}$ via
$q_{\pi} = k_{\pi}L/2\pi$ and $\phi$ is a kinematic function defined in \cite{LLfactor}. Once $q_{\pi}$ is known, $\delta$ can be calculated using the
L\"uscher quantisation condition \cite{Luscher_quant}.
\begin{equation}
 n\pi = \delta(k_{\pi}) + \phi(q_{\pi}).
\end{equation}

$E_{\pi\pi}$ is found by fitting the quotient of correlators $
 C_{\pi\pi}/(C_{\pi})^2 \sim A e^{-\Delta E\, t}
$
to extract $\Delta E= E_{\pi\pi} - 2E_{\pi}$. We then get the two-pion energy by calculating $\Delta E+ 2 E_{\pi}$, where in the case of
$p=0$, $E_{\pi}=m_{\pi}$ while for $p=\sqrt{2}\pi/L$, $E_{\pi}$ is found from a 2 parameter fit to the pion correlation function which also
has $p=\sqrt{2}\pi/L$. This method is preferred to directly extracting $E_{\pi\pi}$ from the two-pion correlator because the quotient $C_{\pi\pi}/(C_{\pi})^2$
cancels some common fluctuations in the numerator and denominator and reduces the error.

\begin{figure}[h]
 \centering
\subfigure[\label{sfig:ps1}Comparison of calculated phase shift with experimental results \cite{Hoogland, Losty, schenk_curve}.]{\includegraphics*[width=0.4\textwidth]{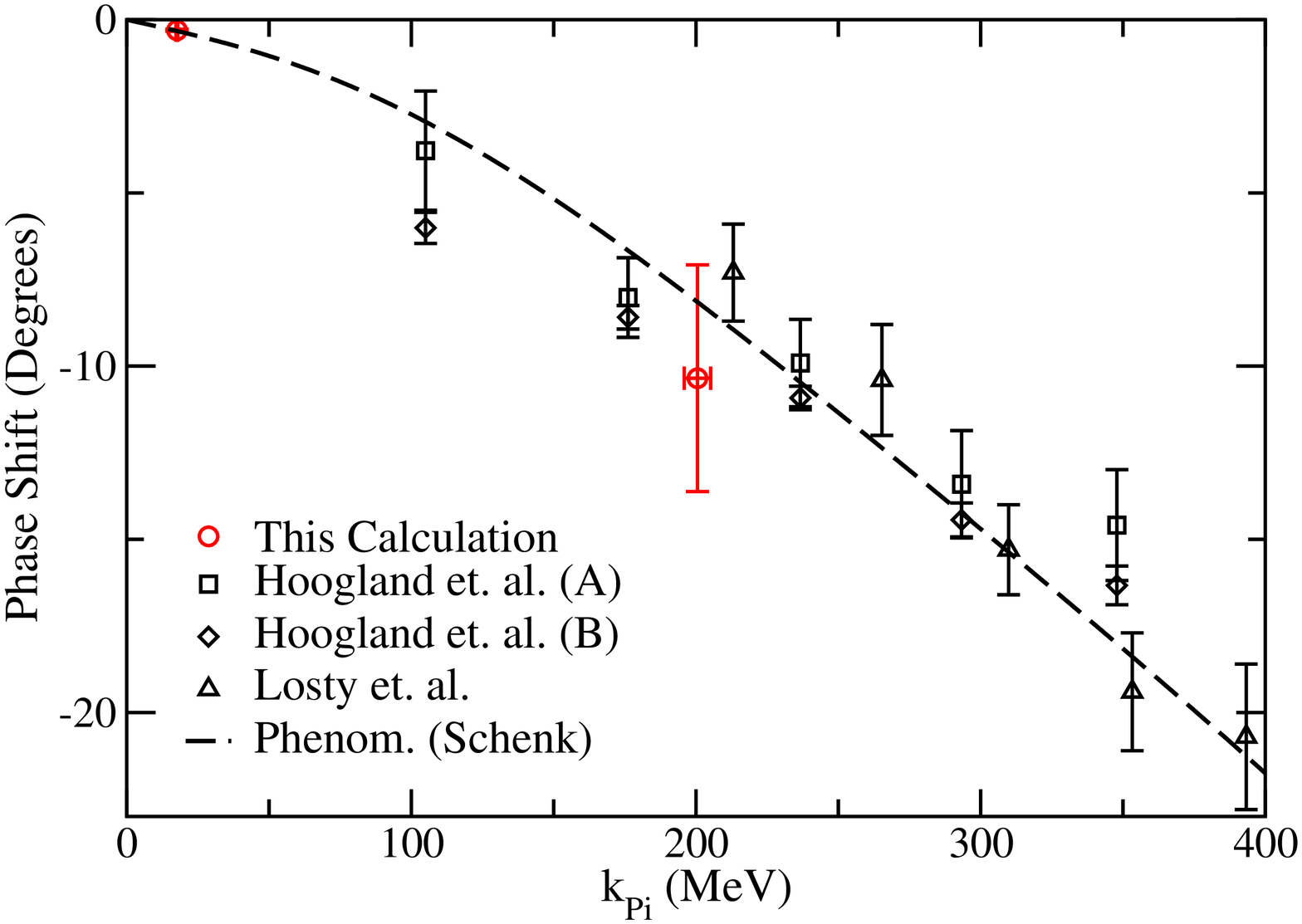}} \qquad
\subfigure[\label{sfig:ps2}Zoom of $2(a)$ showing that for small $k$ $\delta$ is approximately linear. The scattering length is calculated using chiral perturbation theory \cite{Colangelo}.]{\includegraphics*[width=0.4\textwidth]{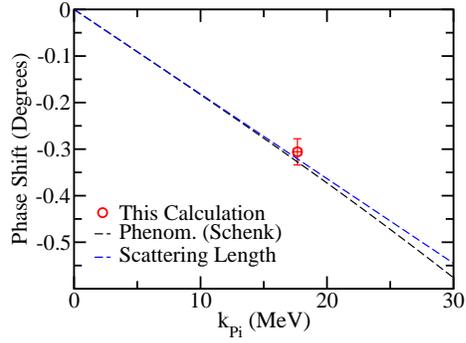}}
\caption{\label{fig:phase_shift}Plot of $I=2$ two-pion s-wave phase shift against momentum $k_{\pi}$. The results from $p=0$ and $p=\sqrt{2}\pi/L$ are shown in red.}
\end{figure}

The pion momentum $k_{\pi}$ in the two-pion state is determined from the two-pion energy using the dispersion relation $E_{\pi\pi} = 2 \sqrt{m^2_{\pi} + k^2_{\pi}}$. 
It differs from $p=0,\, \sqrt{2}\pi/L$ due to interactions between the two pions. Results
for $E_{\pi\pi}$, $k_{\pi}$, $q_{\pi}$ and $\delta$ are presented in table \ref{tab:p_delta}. 
$\partial \phi/\partial q$ can be calculated 
analytically so the only unknown in equation \eqref{eq:amp} is $\partial \delta/\partial q$. 
The results for the phase shift can be plotted against $k_{\pi}$ and compared with experiment \cite{Hoogland, Losty}. This is done in figure \ref{sfig:ps1} and
we see good agreement with experiment. For $p=0$ we make the approximation that $\delta$ is linear with $k_{\pi}$ in order to 
calculate $\partial \delta/ \partial q_{\pi}$ (see figure \ref{sfig:ps2}). For $p=\sqrt{2}\pi/L$ we use the phenomenological curve \cite{schenk_curve} shown in figure 
\ref{sfig:ps1} to calculate the derivative of the phase shift at the corresponding value of $q_{\pi}$. The derivative of the phase shift is
found to be a small factor in comparison with $\partial \phi/ \partial q_{\pi}$.
Results for $\partial \phi/\partial q_{\pi}$ 
and $\partial \delta / \partial q_{\pi}$ are shown in table \ref{tab:derivs}.

\begin{table}[h]
\centering
\caption{Two pion energy and s-wave phase shift \label{tab:p_delta}}
\begin{tabular}{|c|c|c|c|c|}
\hline
$p$ & $E_{\pi\pi}$ (MeV) & $k_{\pi}$ (MeV)& $q_{\pi}$ & $\delta$ (degrees) \\
\hline
0 & 285.9(4.6)& 17.55(61) & 0.0655(21) & -0.306(29)\\
$\sqrt{2}\pi/L$ & 489.2(8.1)& 199.2(3.8) & 0.743(11) & -10.4(3.3)\\
\hline
\end{tabular}
\end{table}

\begin{table}[h]
\centering
\caption{Contributions to Lellouch-L\"uscher factor \label{tab:derivs}}
\begin{tabular}{|c|c|c|c|}
\hline
$p$ & $\partial\phi/\partial q_{\pi}$ & $ \partial \delta / \partial q_{\pi} $& LL \\
\hline
0 & 0.239(14)& -0.0815(50) &0.9636(22) \\
$\sqrt{2}\pi/L$ & 5.039(35)& -0.2927(52)& 0.933(11) \\
\hline
\end{tabular}
\end{table}

The amplitudes $A_i$ are related to the physical decay amplitude $A_2$ via
\begin{equation}
 A_2 = a^{-3} \sqrt{\frac{3}{2}} G_F V_{ud} V^*_{us} \sum_{i,j} C_i(\mu) Z_{ij}(\mu) A_j,
\end{equation}
where $C_i$ are the Wilson Coefficients and $Z_{ij}$ are the renormalization constants, calculated using non-perturbative renormalization (NPR). 
The factor $\sqrt{3/2}$ is needed
to convert from the unphysical $K^+ \rightarrow \pi^+ \pi^+$ amplitudes back to the physical $K^+ \rightarrow \pi^+ \pi^0$ amplitudes. 
At present only $Z_{(27,1)}$ has
been calculated; for the $(8,8)$ and $(8,8)_{\text{mx}}$ operators (which mix under renormalization)
we make the approximation $Z_{ij} = 0.9 Z_q^2\delta_{ij}$. A full calculation
of  $Z_{ij}$ for the $(8,8)$ and $(8,8)_{\text{mx}}$ operators on the lattice is currently under way~\cite{Garron_proceedings}.
Comparing $E_{\pi\pi}$ for $p=\sqrt{2}\pi/L$ with the kaon mass $m_K=507.4(8.5)$ MeV we see that the decay is nearly energy conserving,
so we use the results from $p=\sqrt{2}\pi/L$ to compute $A_2$.
Results for Re$(A_2)$ and Im$(A_2)$ for the four different kaon source times are shown in table \ref{tab:A2}. Our final result for 
Re$(A_2)$ and Im$(A_2)$ is an error weighted average (EWA) over the four kaon source times.

\begin{table}[h]
\centering
\caption{\label{tab:A2}Final results for $A_2$. The errors quoted are statistical errors only.}
\begin{tabular}{|c|c|c|}
\hline
$t_K$& Re$(A_2)$(units of $10^{-8}$ GeV) & Im$(A_2)$(units of $10^{-13}$ GeV) \\
\hline
20 &1.33(11)& -8.11(52)\\
24 &1.44(11)& -8.77(60)\\
28 &1.53(13)& -8.58(58)\\
32 &1.20(16)& -9.01(75)\\
\hline
EWA &1.396(81)& -8.46(45)\\
\hline
\end{tabular}
\end{table}

\vspace{-0.05 in}
\section{Systematic Error}
\vspace{-0.05 in}

The major sources of systematic error in the determination of $A_2$
are scaling violations, finite volume effects, partial quenching,
uncertainty in $\partial\delta/\partial q$, and the fact that the
masses and momentum are slightly different from their physical values.
Furthermore, the approximation made for the renormalization constants for
the (8,8) operators introduces a large systematic error into
$\text{Im}(A_2)$ which we will estimate as 20\%.  $A_2$ is very
sensitive to scaling violations because it is proportional to
$a^{-3}$.  We estimate this systematic by calculating $\text{Re}(A_2)$
with a lattice spacing determined from $f_K$, $m_\Omega$, and $r_0$
respectively, and find a fluctuation of 8.5\% among the three values.
  For finite volume
effects we estimate 7\% for the systematic error using finite volume
chiral perturbation theory for the $K\to\pi\pi$ matrix
elements~\cite{aubin_laiho,soni_laiho}.
One expects that for $\Delta I=3/2$ decays partial quenching will
introduce small errors and in~\cite{Lightman_proceedings_08} the use of partial quenching has been shown to
affect $\text{Re}(A_2)$ by about 2\%.  A value of
$\partial\delta/\partial q$ that is rather larger in magnitude is
obtained just by putting a straight line through the two-pion phase
shift data points from this calculation in figure \ref{sfig:ps1}; this
value yields a result for $\text{Re}(A_2)$ that differs by 2\% which
we use as our conservative estimate of this systematic.  Finally, a
$K\to\pi\pi$ calculation on $24^3$ quenched lattices was done for a
variety of meson masses and two-pion energies~\cite{Lightman_thesis},
and shows that the deviations of these parameters from their physical
values in the present calculation causes an 1.2\% difference in
$\text{Re}(A_2)$.  Adding all errors in quadrature results in a preliminary
estimate of 11\% for the systematic error in $\text{Re}(A_2)$ and 23\%
for the systematic error in $\text{Im}(A_2)$.

\vspace{-0.05 in}
\section{Conclusions}
\vspace{-0.05 in}

We have presented preliminary results for the $\Delta I = 3/2$ $K \rightarrow \pi\pi$ decay amplitude on $32^3$ lattices with $2+1$ flavours
of DWF and the Iwasaki-DSDR gauge action. We find $m_{\pi}=141.9(2.3)$~MeV, $m_K = 507.4(8.5)$~MeV and $E_{\pi\pi} = 489.2(8.1)$~MeV.
The main contribution to Re$(A_2)$ is expected to be from the $(27,1)$ operator, and our result 
$1.396(081)_{\text{stat}}(160)_{\text{sys}}\times 10^{-8}$~GeV can be compared
to the experimental result of 1.5$\times 10^{-8}$ GeV and is found to agree within error. This is the first time a calculation of this type has been achieved.
Im$(A_2)$ is dominated by the operators in the $(8,8)$ representation, 
so we expect there to be a large systematic error on Im($A_2$) due to the approximation made for $Z_{ij}$. This is reflected in our final answer
 Im$(A_2) = -8.46(45)_{\text{stat}}(1.95)_{\text{sys}} \times 10^{-13}$ GeV.
This source of systematic error will be eliminated once the NPR calculation has been completed.

We thank all of our colleagues in the RBC
and UKQCD collaborations for helpful discussions and the development and support of the QCDOC hardware and software infrastructure which was 
essential to this work. In addition we acknowledge Columbia University, RIKEN, BNL, ANL, and the U.S. DOE for providing the facilities on which 
this work was performed. This research used resources of the Argonne Leadership Computing Facility at Argonne National Laboratory, 
which is supported by the Office of Science of the U.S. DOE under contract DE-AC02-06CH11357. 
This work was supported in part by U.S. DOE grant number DE-FG02-92ER40699. Elaine Goode is supported by an
STFC studentship and grant ST/G000557/1 and by EU contract MRTN-CT-2006-03542 (Flavianet).

\end{document}